\documentclass[aps,prx,reprint,showkeys,amsfonts,amssymb,amsmath,showpacs,floatfix,fleqn]{revtex4-1}

\usepackage{graphicx,color,psfrag,easybmat,ulem,multirow}

\renewcommand{\exp}[1]{e^{#1}}
\renewcommand{\vec}[1]{\mathbf{#1}}
\newcommand{\mat}[1]{\mathbf{#1}}
\newcommand{\order}{\mathcal{O}}
\newcommand{\degree}{^{\circ}}

\begin{document}
\title{Instabilities and Patterns in Coupled Reaction-Diffusion Layers}
\author{Anne J. Catll{\'a}}
\email{catllaaj@wofford.edu}
\homepage{\url{http://webs.wofford.edu/catllaaj}}
\affiliation{Dept. of Mathematics, Wofford College Spartanburg, SC 29303, USA}
\author{Amelia McNamara}
\altaffiliation[Currently at: ]{Dept. of Statistics, UCLA, Los Angeles, CA 90024.}
\author{Chad M. Topaz}
%\email{ctopaz@macalester.edu}
%\homepage{\url{http://www.macalester.edu/~ctopaz}}
\affiliation{Dept. of Mathematics, Statistics, and Computer Science, Macalester College, St. Paul, MN 55105, USA}
\date{\today}

\begin{abstract}
We study instabilities and pattern formation in reaction-diffusion layers that are diffusively coupled. For two-layer systems of identical two-component reactions, we analyze the stability of homogeneous steady states by exploiting the block symmetric structure of the linear problem. There are eight possible primary bifurcation scenarios, including a Turing-Turing bifurcation that involves two disparate length scales whose ratio may be tuned via the inter-layer coupling.  For systems of $n$-component layers and non-identical layers, the linear problem's block form allows approximate decomposition into lower-dimensional linear problems if the coupling is sufficiently weak. As an example, we apply these results to a two-layer Brusselator system. The competing length scales engineered within the linear problem are readily apparent in numerical simulations of the full system. Selecting a $\sqrt{2}$:1 length scale ratio produces an unusual steady square pattern.
\end{abstract}


\maketitle

\section{Introduction}
\label{sec:introduction}

In 1952, Alan Turing hypothesized that reaction and diffusion could compete to create stationary spatial patterns \cite{Tur1952}. This hypothetical mechanism for biological morphogenesis has been the theoretical foundation for decades of work on Turing patterns, which form when a rapidly diffusing activator interacts with a slowly diffusing inhibitor.  Nearly 40 years later, experimentalists observed these patterns in a chemical reaction-diffusion system~\cite{CasDulBoi1990}.  Since then, chemical systems have been the canonical testing ground for Turing patterns.  

A variation on the classic Turing system is the multi-layered system, in which each layer is a reaction-diffusion system that is diffusively coupled to adjacent layers. These coupled systems are common in the biological world, seen in neural, developmental, and ecological contexts~\cite{EpsBerDol2008}.  One example from neuroscience is a neural-glial network, consisting of a layer of neurons connected diffusively to a layer of glial cells, where each layer exhibits dynamics at different time scales.  The chemicals released at a tripartite synapse (one glial cell and a pair of neurons) and their effect on those cells are known~\cite{AraParSan1999,NadJun2004}; however, the effect of glial cells on the network level remains a subject of ongoing study~\cite{HalHay2010}.  Understanding how coupled layers influence one another contributes to our understanding of these networks.

Though experimental studies of the biological systems are quite difficult, investigations of the fundamental properties of coupled reaction-diffusion systems have progressed via chemical experiments. Experimentalists employ two thin gels (which contain the reactants) that are put in contact with one another.  By adding or removing a permeable membrane between the layers and by adjusting its properties, the coupling strength can be altered.  This approach with the chlorine dioxide - iodine - malonic acid (CDIMA) reaction has produced superlattice patterns called black-eyed and white-eyed patterns which involve wavelength ratios of nearly 2:1; other ratios were not feasible for this reaction and experimental configuration~\cite{BerDolYan2004}. Recent experiments have exploited the photosensitivity of the CDIMA chemical reaction, using an external light source to probe the interaction between different forced patterns \cite{MigDolEps2011}. For a broad overview of experimental and numerical results for some multi-layer systems, see~\cite{EpsBerDol2008}. 

A few theoretical studies of multilayer systems have taken place in the setting of diffusively coupled ordinary differential equations; this framework neglects spatial dependence within layers (and hence, spatiotemporal pattern formation) but is more easily analyzed than the spatial case. Linear stability analysis and numerical bifurcation studies reveal regimes of in-phase and out-of-phase oscillations of coupled Brusselators~\cite{VolRom1995}, as well as regimes of synchronization and chaos in coupled Oregonators~\cite{LiuWenLee1997}.  For Brusselators, regions of in-phase waves and echo waves, whose phase differs by half the period, can be determined analytically~\cite{ZhoZha2001,ZhoZha2002}. 

Work incorporating spatial dependence within layers has also used linear stability and bifurcation analyses to determine and understand possible patterns, now in the setting of partial differential equation models of chemical reactions. For coupled Oregonators, simulations reveal twinkling eye patterns, Turing spots arranged in a hexagonal lattice that oscillate 120 degrees out of phase with their nearest neighbors, and traveling waves in Turing structures, such as pinwheels in spots and traveling waves in labyrinths~\cite{YanEps2003}.  A numerically computed dispersion relation suggests that the twinkling eye pattern is due to an interaction of Turing and Hopf modes, whereas the traveling wave patterns are formed via a short wave instability \cite{YanEps2003}.  Similar analyses have also elucidated the bifurcations to time-dependent Turing states and superlattices in coupled Lengyel-Epstein equations as parameters vary~\cite{YanEps2004}, in the presence of a delay~\cite{JiLi2005}, and with external forcing~\cite{JiLi2006}.  Simulations of coupled Brusselators demonstrate superposition patterns, twinkling eye patterns, and black-eyed and white-eyed superlattices.   They are the result of two interacting Turing modes and occur when the ratio of the interacting modes is close to $\sqrt{3}$:1, 2:1, 3:1,~\cite{YanDolZha2002} or 4:1~\cite{LiuPan2010}. The Jacobian matrix of this system has been studied numerically to understand these patterns~\cite{KytKasBar2007,LiuPan2010}. One analytical study of coupled Brusselators used a linear stability analysis to obtain conditions for existence of steady states and non-constant solutions~\cite{ZhoMu2010}.  Extending work on coupled layers,~\cite{NakMik2010} studies networks of reaction-diffusion systems, which are closely related to the BZ-AOT experimental system~\cite{VanEps2003}.   

Analytical calculations for layered reaction-diffusion systems can be difficult because of dimensionality. For instance, with $m$ layers of $n$-component reaction-diffusion systems, the linear problem is $mn \times mn$; this suggests why even the linear results for the papers referenced above are largely numerical. In this paper, we show how the linear calculations may be simplified and harnessed to engineer certain aspects of nonlinear pattern formation. For the case of a two-layer, two-component system, we exploit the block symmetric form of the Jacobian to analyze the stability of homogeneous steady states. There are eight possible primary bifurcation scenarios, and we determine conditions under which each occurs. One possibility is a Turing-Turing bifurcation that involves two disparate length scales whose ratio may be tuned via the inter-layer coupling.  For systems of $n$-component layers and non-identical layers, the linear problem's block form allows approximate decomposition into lower-dimensional linear problems if the coupling is sufficiently weak. We apply some results to a two-layer Brusselator system near the Turing-Turing bifurcation. The competing length scales engineered within the linear problem are readily apparent in numerical simulations of the full system. Selecting a $\sqrt{2}$:1 ratio produces a steady square pattern.  Square superlattice Turing patterns have been previously reported, initially in \cite{YanDolZha2006}, under the influence of external forcing.  However, to our knowledge, a steady pattern of simple Turing squares (moreover, one obtained without forcing) has not been previously reported.

The rest of this paper is organized as follows. Sec.~\ref{sec:linear1} presents a linear stability analysis of coupled layers of reaction-diffusion systems, describing in detail the primary bifurcations for the case of two-layer, two-component systems. The linear algebra necessary to simplify the calculations is developed in the Appendices. Sec. \ref{sec:makingpatterns} applies some of the results in order to engineer nonlinear patterns containing a desired length scale ratio, as demonstrated in simulations. We conclude in Sec.~\ref{sec:conclusion}.

\section{Linear analysis of coupled reaction-diffusion layers}
\label{sec:linear1}

We now present results for the (in)stability of trivial states of coupled reaction-diffusion layers. In Sec. \ref{sec:linearderiv} through \ref{sec:primbif} we focus on two-layer systems of identical two-component layers. Exploiting the block symmetric structure of the linearized problem, we find convenient expressions for the eigenvalues that are easily analyzed, and we enumerate the possible primary bifurcations. In Sec. \ref{sec:extensions}, we mention a few brief results applying to systems with non-identical layers, more complicated coupling schemes, and systems with more chemical components. Because we begin with generic reaction-diffusion equations, the results are readily applied to specific systems such as the Brusselator \cite{PriLef1968}, the Lengyel-Epstein model \cite{LenEps1991}, and so forth.

\subsection{Derivation of linearized problem}
\label{sec:linearderiv}

We begin with nonlinear equations describing identical two-component reaction-diffusion layers that are coupled together,
\begin{subequations}
\label{eq:twolayer}
\begin{alignat}{7}
\dot{U}_i&=&\alpha(U_j-U_i)&+F(U_i,V_i)&+\phantom{D}\nabla^2U_i,\\
\dot{V}_i&=&\beta (V_j-V_i)&+G(U_i,V_i)&+D\nabla^2V_i.
\end{alignat}
\end{subequations}
This model describes layers that are identical in their underlying chemical and physical properties.  Throughout this section, $i,j=1,2$, $i\neq j$ indicates the layer. $U_i(\vec{x},t), V_i(\vec{x},t)$ are chemical concentration fields, $\vec{x}$ is the spatial coordinate, $t$ is time, and the over dot represents a time derivative. The functions $F$ and $G$ are reaction kinetics terms whose functional form depends on the particular chemical model under consideration. The diffusivity of $U$ is set to unity by a rescaling of the spatial coordinate; the diffusivity of $V$ is $D$. Without loss of generality, assume $V$ to be the more rapidly diffusing species, so that $D > 1$. Finally,  $\alpha, \beta \geq 0$ are coefficients of diffusive coupling between the systems.  

We wish to study bifurcations from a spatially uniform steady state. As pointed out in \cite{YanEps2004}, different types of uniform states may be possible. One possibility is that the concentrations of the two layers are identical. A second possibility is that the two layers have distinct (uniform) concentrations even though the underlying equations are the same. In practice, the types of steady states that exist are determined by the particular form of the reaction kinetics functions $F,G$ and the chemical parameters therein. In this section, as the simplest case, assume that the two layers share the same uniform steady state. We relax this assumption in Sec. \ref{sec:extensions}.

Let the uniform steady state be $U_i=U^*$, $V_i=V^*$. Write the chemical fields as a perturbation around the steady state, and express the perturbation as a superposition of Fourier modes.
\begin{equation}
\begin{pmatrix} U_i \\ V_i \end{pmatrix} = \begin{pmatrix} U^* \\ V^* \end{pmatrix} + \sum_\vec{q} \begin{pmatrix} u_{i,\vec{q}} \\ v_{i,\vec{q}} \end{pmatrix} \exp{i \vec{q} \cdot \vec{x}}.
\end{equation}
The perturbation has wave number $q = |\vec{q}|$ and $u_{i,\vec{q}}(t)$ and $v_{i,\vec{q}}(t)$ are Fourier wave amplitudes. The summation and the admissible $\vec{q}$ must be interpreted in a manner consistent with the boundary conditions of the governing equations; for instance, if the equations are posed on an unbounded domain, then the summation is actually an integral over all $\vec{q}$ (per the Fourier transform). To assess the stability of the steady state, study the linearized problem governing the perturbations,
\begin{subequations}
\label{eq:linear1}
\begin{alignat}{9}
\dot{u}_i&=&\alpha(u_j - u_i) &+ au_i &+ bv_i &- \phantom{D}q^2 u_i,\\
\dot{v}_i&=&\beta(v_j - v_i) &+ cu_i &+ dv_i &- Dq^2 v_i.
\end{alignat}
\end{subequations}
For brevity, and as a convenient abuse of notation, we have suppressed the $q$ dependence in the subscript of the Fourier wave amplitudes. The coefficients $a$, $b$, $c$, and $d$ are given by
\begin{subequations}
\begin{alignat}{9}
a & = & \frac{\partial F}{\partial U}\Bigr|_{(U^*,V^*)}, \qquad &b&=& \frac{\partial F}{\partial V}\Bigr|_{(U^*,V^*)},\\
c & = & \frac{\partial G}{\partial U}\Bigr|_{(U^*,V^*)}, \qquad &d&=& \frac{\partial G}{\partial V}\Bigr|_{(U^*,V^*)}.
\end{alignat}
\end{subequations}

It is convenient to write the problem in matrix form. Let $\vec{u} = (u_1,v_1,u_2,v_2)^T$. The linearized problem is
\begin{equation}
\label{eq:ourlinear}
\dot{\vec{u}} = \mat{L} \vec{u}, \quad 
\mat{L}=\begin{pmatrix} \mat{P} & \mat{Q} \\ \mat{Q} & \mat{P} \end{pmatrix}.
\end{equation}
$\mat{L}$ is of block symmetric form, with blocks
\begin{subequations}
\begin{eqnarray}
\mat{P} & = & \begin{pmatrix} a - q^2 - \alpha & b \\ c & d-Dq^2-\beta \end{pmatrix},\\
\mat{Q} & = & \begin{pmatrix} \alpha & 0 \\ 0 & \beta \end{pmatrix}.
\end{eqnarray}
\end{subequations}

We show in Appendix \ref{sec:block} that the eigenvalues of $\mat{L}$ are the eigenvalues of $\mat{L}_1 = \mat{P} + \mat{Q}$ and $\mat{L}_2 = \mat{P} - \mat{Q}$. Hence, the linear problem decomposes conveniently into two subproblems described by the matrices
\begin{subequations}
\label{eq:linearops}
\begin{eqnarray}
\mat{L}_1 & = & \begin{pmatrix} a - q^2 & b \\ c & d-Dq^2 \end{pmatrix},\\
\mat{L}_2 & = & \begin{pmatrix} a - q^2 - 2 \alpha & b \\ c & d-Dq^2 -2\beta \end{pmatrix}.
\end{eqnarray}
\end{subequations}
The matrix $\mat{L}_1$ is simply the Jacobian corresponding to a solitary reaction diffusion layer. The effect of the coupling between layers is seen via $\mat{L}_2$. Though the full linear problem is four-by-four with a quartic characteristic polynomial, the problem decomposes into these two $2 \times 2$ problems, facilitating analysis.

\subsection{Global extrema of trace and determinant}

\label{sec:globalextrema}

In Sec. \ref{sec:primbif} we will consider different bifurcation scenarios by analyzing the trace $\tau_{1,2}(q)$ and determinant $\Delta_{1,2}(q)$ of $\mat{L}_{1,2}$,
\begin{subequations}
\label{eq:tracedet}
\begin{align}
\tau_1(q) &= a + d - (D+1)q^2, \\
\tau_2(q) &= \tau_1 - 2(\alpha+\beta), \\
\Delta_1(q) & = D q^4 - (aD+d)q^2 + ad-bc, \\
\Delta_2(q) & =  \Delta_1 + 2(\alpha D+\beta)q^2 \\
& \phantom{=\ }+2(-\alpha d - \beta a + 2\alpha\beta). \nonumber
\end{align}
\end{subequations}
Here, we present two helpful observations.

First, $\tau_{1,2}(q)$ are quadratic in $q$, each with a negative leading coefficient and no $q^1$ term, and hence have global maxima at $q = 0$. We have
\begin{subequations}
\label{eq:tauconds}
\begin{eqnarray}
\tau_1(0) & = & a+d,\\
\tau_2(0) & = & a + d -2(\alpha+\beta) \leq \tau_1(0).
\end{eqnarray}
\end{subequations}

Second, $\Delta_{1,2}(q)$ are even-powered quartics, each with a positive leading coefficient. Thus, these quantities have global minima. Label them $(q_{1,min},\Delta_{1,min})$ and $(q_{2,min},\Delta_{2,min})$. Whether the global minima occur at zero or nonzero $q$ depends on the sign of the quadratic coefficient. For $\Delta_1(q)$,
\begin{subequations}
\label{eq:q11}
\begin{gather}
\textrm{If $aD+d > 0$:}\nonumber \\
q_{1,min}^2 = \frac{aD+d}{2D}, \label{eq:q1min} \\
\Delta_{1,min} = -\frac{(aD-d)^2+4Dbc}{4D}, \label{eq:delta1min}
\end{gather}
\end{subequations}
or
\begin{subequations}
\label{eq:q12}
\begin{gather}
\textrm{If $aD+d \leq 0$:}\nonumber \\
q_{1,min}^2 = 0, \\
\Delta_{1,min} = ad-bc.
\end{gather}
\end{subequations}
Similarly, for $\Delta_2(q)$,
\begin{subequations}
\label{eq:q21}
\begin{gather}
\textrm{If $aD+d - 2\alpha D - 2\beta > 0$:}\nonumber \\
q_{2,min}^2 = \frac{aD+d-2\alpha D - 2\beta}{2D}, \label{eq:q2min} \\
\Delta_{2,min} = -\frac{(aD-d-2\alpha D + 2\beta)^2+4Dbc}{4D}, \label{eq:delta2min}
\end{gather}
\end{subequations}
or
\begin{subequations}
\label{eq:q22}
\begin{gather}
\textrm{If $aD+d - 2\alpha D - 2\beta \leq 0$:}\nonumber \\
q_{2,min}^2 = 0, \\
\Delta_{2,min} = ad-bc-2\alpha d - 2 a \beta+4\alpha\beta.
\end{gather}
\end{subequations}
Finally, note that $q_{2,min} \leq q_{1,min}$ since $\alpha,\beta \geq 0$.

\subsection{Primary bifurcations}
\label{sec:primbif}

\begingroup
%\squeezetable
\begin{table*}
\caption{Summary of possible primary bifurcations of the homogeneous steady state of (\ref{eq:twolayer}). The four-dimensional linearized problem consists of two two-dimensional  sub-problems per (\ref{eq:linearops}). We distinguish between two different classes of bifurcations. First, there are bifurcations due to eigenvalues in $\mat{L}_1$, which also occur in single-layer (traditional) two-component reaction-diffusion systems. These bifurcations are captured in Cases I - IV and are very well-known. Second, there are bifurcations due to eigenvalues in $\mat{L}_2$, and thus which depend on the diffusive coupling between the two layers. These are cases V - VIII. Below, a dash indicates no bifurcation, H indicates Hopf, T indicates Turing, and TH indicates Turing-Hopf. For each scenario, we state generic conditions on the traces and determinants $\tau_{1,2}(q)$ and $\Delta_{1,2}(q)$ in (\ref{eq:tracedet}). In practice, we enforce these conditions by controlling the global extrema of $\tau_{1,2}(q)$ and $\Delta_{1,2}(q)$; see Sec. \ref{sec:primbif} for details. In the table, the wave number $q_c$ refers to a critical wave number; cases VII and VIII have two critical wave numbers. \label{tab:bifurcations}}
\begin{ruledtabular}
\begin{tabular}{ccccccc}
\multirow{2}{*}{\textbf{Case}} & \multicolumn{2}{l}{\textbf{Bifurcation due to}} & \multirow{2}{*}{$\tau_1(q)$} & \multirow{2}{*}{$\tau_2(q)$} & \multirow{2}{*}{$\Delta_1(q)$} & \multirow{2}{*}{$\Delta_2(q)$} \\  
& $\mat{L}_1$ & $\mat{L}_2$ & & & &\\  \hline
&  & & & & & \\
I & - & - & $\tau_1(q) < 0$ & $\tau_2(q) < 0$ & $\Delta_1(q) > 0$ & $\Delta_2(q) > 0$ \\
& & & & & & \\
II & H & - & $\tau_1(0) = 0$ & $\tau_2(q) < 0$ & $\Delta_1(q) > 0$ & $\Delta_2(q) > 0$ \\
& & &  $\tau_1(q\neq0) < 0$ & & &\\
& & & & & & \\
III & T & - & $\tau_1(q) <0$ & $\tau_2(q) <0$ & $\Delta_1(q_c) = 0$ & $\Delta_2(q)>0$\\
& & & & & $\Delta_1(q\neq q_c)>0$ & \\
& & & & & & \\
IV & TH & - & $\tau_1(0) = 0$ & $\tau_2(q) < 0$ & $\Delta_1(q_c) = 0$ & $\Delta_2(q)>0$\\
& & &  $\tau_1(q\neq0) < 0$ & & $\Delta_1(q\neq q_c)>0$ & \\
& & & & & & \\
V &  - & T & $\tau_1(q) <0$ & $\tau_2(q) <0$ & $\Delta_1(q)>0$ & $\Delta_2(q_c) = 0$ \\
& & & & & & $\Delta_2(q\neq q_c)>0$ \\
& & & & & & \\
VI & H & T & $\tau_1(0) = 0$ & $\tau_2(q) < 0$ & $\Delta_1(q)>0$ & $\Delta_2(q_c) = 0$ \\
& & &  $\tau_1(q\neq0) < 0$ & & &$\Delta_2(q\neq q_c)>0$\\
& & & & & & \\
VII & T & T & $\tau_1(0) < 0$ & $\tau_2(q) < 0$ & $\Delta_1(q_{1,c})=0$ & $\Delta_2(q_{2,c}) = 0$ \\
& & &  & & $\Delta_2(q\neq q_{1,c})>0$ & $\Delta_2(q\neq q_{2,c})>0$\\
& & & & & & \\
VIII & TH & T & $\tau_1(0) = 0$ & $\tau_2(q) < 0$ & $\Delta_1(q_{1,c})=0$ & $\Delta_2(q_{2,c}) = 0$ \\
& & & $\tau_1(q\neq 0)<0$ & & $\Delta_2(q\neq q_{1,c})>0$ & $\Delta_2(q\neq q_{2,c})>0$\\

\end{tabular}
\end{ruledtabular}
\end{table*}
\endgroup

We now consider possible primary bifurcation scenarios. Naively, $\mat{L}_1$ and $\mat{L}_2$ may each give rise to four different primary bifurcation scenarios: none (linear stability), Hopf bifurcation (H), Turing bifurcation (T), and Turing-Hopf bifurcation (TH). Since the full linear problem comprises $\mat{L}_{1,2}$, there would be $4 \times 4 = 16$ primary bifurcation scenarios.

However, due to the particular form of $\mat{L}_{1,2}$, any scenario involving a primary Hopf bifurcation in $\mat{L}_2$ (that is, H or TH) is impossible. To see this, assume a primary Hopf bifurcation due to $\mat{L}_2$. This requires $\tau_2(0)=0$ per (\ref{eq:tauconds}). However, since $\tau_2(0) \leq \tau_1(0)$ (with equality achieved only in the trivial case $\alpha=\beta=0$), the assumption means that a Hopf bifurcation would already have occurred due to $\mat{L}_1$, and hence the assumed bifurcation due to $\mat{L}_2$ would not, in fact, be the primary one. Therefore, all bifurcation scenarios involving primary H or TH bifurcations due to $\mat{L}_2$ are prohibited. This eliminates eight of the 16 possible scenarios. Of course, if the layers were not identical, this result would not hold, and other primary bifurcations might be possible. For an example involving different Hopf bifurcations, see the nonspatial two-cell chemical model in \cite{BouEleArn1987}.

The remaining eight possible primary bifurcation scenarios are enumerated in Table \ref{tab:bifurcations}. We find the conditions for each case by analyzing $\tau_{1,2}(q)$ and $\Delta_{1,2}(q)$ in the usual way to determine when a single eigenvalue or pair of eigenvalues crosses the imaginary axis, with all other eigenvalues contained in the left half of the complex plane.  In these cases we distinguish between two different classes of bifurcations. First, there are bifurcations due to eigenvalues in $\mat{L}_1$, which also occur in single-layer (traditional) two-component reaction-diffusion systems. These bifurcations are captured in \mbox{Cases I - IV} and are very well-known. Second, there are bifurcations due to eigenvalues in $\mat{L}_2$, and thus which depend on the diffusive coupling between the two layers. These are  \mbox{Cases V - VIII}. They correspond to \mbox{Cases I - IV} but with an additional Turing bifurcation due to~$\mat{L}_2$.

We apply the generic conditions in Table \ref{tab:bifurcations} to our specific linear problem (\ref{eq:ourlinear}) by enforcing conditions on the global extrema of $\tau_{1,2}(q)$, $\Delta_{1,2}(q)$.  First, focus on the trace (the fourth and fifth columns of Table \ref{tab:bifurcations}.) An examination of~(\ref{eq:tracedet}) and~(\ref{eq:tauconds}) shows that if $\tau_1(0)<0$, then $\tau_1(q\neq0)<0$, and similarly for $\tau_2(q)$.  Recall also, as noted in~(\ref{eq:tauconds}), that $\tau_2(0)\leq\tau_1(0)$.   Thus for our model, the condition that $\tau_2(q)<0$, required for all of the bifurcations in Table~\ref{tab:bifurcations}, is subsumed in the condition on $\tau_1(0)$ and $\tau_1(q\neq 0)$ in that table. 

Now focus on conditions for the determinants (the sixth and seventh columns of Table \ref{tab:bifurcations}). These are easily enforced by controlling $\Delta_{1,min}$ and $\Delta_{2,min}$ as given by (\ref{eq:q11}) - (\ref{eq:q22}). In cases of Turing bifurcations, the critical wave numbers $q_{1,c}$ and/or $q_{2,c}$ are identified with the locations of the global minima, namely $q_{1,min}$ and/or $q_{2,min}$. 

There is still the matter of which expressions out of (\ref{eq:q11}) - (\ref{eq:q22}) apply for each case. For Cases I and II, either (\ref{eq:q21}) or (\ref{eq:q22}) will apply for $(q_{2,min},\Delta_{2,min})$, depending on chemical kinetics and parameters. If (\ref{eq:q21}) applies, then (\ref{eq:q11}) must apply for $(q_{1,min},\Delta_{1,min})$ since $q_{2,min} \leq q_{1,min}$. If (\ref{eq:q22}) applies, then one of (\ref{eq:q11}) or (\ref{eq:q12}) will apply for $(q_{1,min},\Delta_{1,min})$, depending on chemical kinetics and parameters.  Since a Turing bifurcation occurs at a nonzero wave number, (\ref{eq:q11}) applies for $\Delta_{1,min}$ in Cases III and IV. In these cases, either (\ref{eq:q21}) or (\ref{eq:q22}) might apply for $\Delta_{2,min}$, depending on chemical kinetics and parameters.  Similarly, in Cases V - VIII, (\ref{eq:q21}) applies for $\Delta_{2,min}$. Since $q_{2,min} \leq q_{1,min}$, (\ref{eq:q11}) applies for $\Delta_{1,min}$.

\subsection{Extensions to other layered reaction-diffusion systems}
\label{sec:extensions}

Suppose each layer comprises a reaction-diffusion system with $n$ chemical components. Then generalizing (\ref{eq:twolayer}), the governing equations are
\begin{equation}
\dot{\vec{U}}_i = \mat{Q} (\vec{U}_j - \vec{U}_i) + \vec{F}(\vec{U}_i) + \mat{D} \nabla^2 \vec{U}_i.
\end{equation}
As before, $i,j$ = 1,2. $i \neq j$ indicates the layer. $\vec{U}_i(\vec{x},t) \in \mathbb{R}^n$ is a vector containing concentrations of the $n$ chemical components in layer $i$. The vector function $\vec{F} \in \mathbb{R}^n$ describes reaction kinetics. The $n \times n$ diagonal matrix $\mat{Q}$ contains coupling coefficients,
\begin{equation}
\mat{Q} = \begin{pmatrix}
\alpha_1 & &  \multicolumn{3}{c}{\text{\kern0.5em\smash{\raisebox{-2ex}{\huge 0}}}} \\
 & \ddots \\
 &  & \alpha_k & \\
 & & & \ddots  \\
\multicolumn{3}{c}{\text{\kern-0.5em\smash{\raisebox{1.5ex}{\huge 0}}}} & & \alpha_n
\end{pmatrix},
\end{equation}
and the $n \times n$ diagonal matrix $\mat{D}$ contains diffusion coefficients,
\begin{equation}
\mat{D} = \begin{pmatrix}
D_1 & &  \multicolumn{3}{c}{\text{\kern0.5em\smash{\raisebox{-2ex}{\huge 0}}}} \\
 & \ddots \\
 &  & D_k & \\
 & & & \ddots  \\
\multicolumn{3}{c}{\text{\kern-0.5em\smash{\raisebox{1.5ex}{\huge 0}}}} & & D_n,
\end{pmatrix},
\end{equation}
and $\nabla^2$ is understood to operate on each element of $\vec{U}_i$.

Assume identical uniform steady states in each layer, $\vec{U}_i = \vec{U}^*$. Then the linearized problem has the block structure (\ref{eq:ourlinear}), as in Sec. \ref{sec:linearderiv}, only now
\begin{equation}
\mat{P} = \displaystyle{\mat{dF}\Bigr|_{\vec{U}^*}} - q^2 \mat{D} - \mat{Q},
\end{equation}
and $\mat{dF}$ is the Jacobian of $\mat{F}$. Of course, now $\mat{P}$ and $\mat{Q}$ are $n \times n$ matrices. Nonetheless, many features are preserved from the $2 \times 2$ case. The eigenvalues of the two-layer system still decompose into the eigenvalues of 
\begin{align}
\mat{L}_1 = \mat{P} + \mat{Q} &=  \displaystyle{\mat{dF}\Bigr|_{\vec{U}^*}} - q^2 \mat{D},\\
\mat{L}_2 = \mat{P} - \mat{Q}  &=  \displaystyle{\mat{dF}\Bigr|_{\vec{U}^*}} - q^2 \mat{D} - 2\mat{Q},
\end{align}
where $\mat{L}_1$ is simply the linear operator corresponding to a single (uncoupled) layer, and $\mat{L}_2$ incorporates the effect of the coupling.

Now, as in \cite{YanEps2004}, allow the uniform steady state to comprise different concentrations in each layer (even though the chemical parameters for each layer are identical) so that
\begin{equation}
\vec{U}_1 = \vec{U}_1^*, \quad \vec{U}_2 = \vec{U}_2^*.
\end{equation}
Then the linearized problem is
\begin{equation}
\label{eq:ourlinear2}
\dot{\vec{u}} = \mat{L} \vec{u}, \quad 
\mat{L}=\begin{pmatrix} \mat{P} & \mat{Q} \\ \mat{Q} & \mat{S} \end{pmatrix},
\end{equation}
where
\begin{subequations}
\begin{eqnarray}
\mat{P} & = & \displaystyle{\mat{dF}\Bigr|_{\vec{U}_1^*}} - q^2 \mat{D} - \mat{Q},\\
\mat{S} & = & \displaystyle{\mat{dF}\Bigr|_{\vec{U}_2^*}} - q^2 \mat{D} - \mat{Q}.
\end{eqnarray}
\end{subequations}
In this case, no simple formula exists for the eigenvalues of $\mat{L}$ in term of $\mat{P}$, $\mat{Q}$, and $\mat{S}$. However, if we assume that coupling is weak, that is $\mat{Q} \to \epsilon \mat{Q}$ where $\epsilon \ll 1$ then the eigenvalues of $\mat{L}$ are approximately equal to the eigenvalues of $\mat{P}$ and the eigenvalues of $\mat{S}$. We show this in Appendix \ref{sec:block2}.

We may also suppose that the two layers have distinct chemical kinetics. For instance, the system might be composed of two coupled Brusselators, but with a different set of chemical control parameters selected for each layer. The governing equations for this case are
\begin{subequations}
\begin{eqnarray}
\dot{\vec{U}}_1 & = & \mat{Q} (\vec{U}_2 - \vec{U}_1) + \vec{F}_1(\vec{U}_1) + \mat{D}_1 \nabla^2 \vec{U}_1,\\
\dot{\vec{U}}_2 & = & \mat{Q} (\vec{U}_1 - \vec{U}_2) + \vec{F}_2(\vec{U}_2) + \mat{D}_2 \nabla^2 \vec{U}_2.
\end{eqnarray}
\end{subequations}
The two distinct chemical kinetics functions $\vec{F}_{1,2}$ and the two distinct matrices of diffusion coefficients $\mat{D}_{1,2}$ reflect the different chemical parameters in each layer. The linearized problem has the same form (\ref{eq:ourlinear2}), only now
\begin{subequations}
\begin{eqnarray}
\mat{P} & = & \displaystyle{\mat{dF}_1 \Bigr|_{\vec{U}_1^*}} - q^2 \mat{D}_1 - \mat{Q},\\
\mat{S} & = & \displaystyle{\mat{dF}_2 \Bigr|_{\vec{U}_2^*}} - q^2 \mat{D}_2 - \mat{Q}.
\end{eqnarray}
\end{subequations}
The results of the previous paragraph still hold. For weak coupling, the eigenvalues are approximately those of $\mat{P}$ and $\mat{S}$.

\section{Multiple length scale selection}
\label{sec:makingpatterns}

Sec. \ref{sec:linear1} showed that the uniform steady state of two identical, coupled reaction-diffusion layers may lose stability via a codimension-two Turing-Turing bifurcation that involves two disparate wave numbers. We now examine this bifurcation in more depth, and explore how the strength of coupling between the layers may be used to tune pattern selection and encourage the formation of spatial patterns with a desired length scale ratio. We apply results to the Brusselator in order to compute length scale ratios as a function of inter-layer coupling strength. Finally, we show via numerical simulation that we are able to engineer nonlinear patterns with pre-selected length scale ratios; this includes a steady square pattern.

\subsection{Length scale ratios}
\label{sec:resonanttriads}

We now focus on Case VII in Table \ref{tab:bifurcations}, which describes the codimension-two Turing-Turing bifurcation.  Our goal is to derive conditions for the Turing-Turing bifurcation in terms of the parameters $a,b,c,d,D,\alpha$, and $\beta$, and to calculate the length scale ratio in terms of these parameters.  Recall that the critical wave numbers for a Turing-Turing bifurcation are $q_{1,c}=q_{1,min}$ and $q_{2,c}=q_{2,min}$ as given by (\ref{eq:q11}) and (\ref{eq:q21}).

The condition $\Delta_{1,min}=0$ enforces a relationship between $a$, $b$, $c$, $d$, and $D$, independent of the coupling parameters $\alpha$ and $\beta$. The condition $\tau_1(0) < 0$ means that $a+d < 0$. Therefore, $a$ and $d$ are oppositely signed. Recalling that (\ref{eq:q11}) applies for Case VII, we know that $aD+d>0$. In order for $q_{1,c}^2$ to be positive,  $a$ must be positive since $D>1$. Hence, $d<0$. For the remainder of this section, we assume that parameters satisfy these inequalities,
\begin{equation}
a>0, \quad d < 0, \quad aD+d > 0. \label{eq:turingconds}
\end{equation}

The next condition in Case VII is $\Delta_{2,min} = 0$. Using (\ref{eq:delta1min}) and substituting (\ref{eq:delta2min}) yields
\begin{equation}
(aD-d)^2 = (aD - d - 2\alpha D + 2\beta)^2,
\end{equation}
from which two possibilities follow. Either
\begin{equation}
\label{eq:bicritcond2}
\beta = \alpha D - aD + d.
\end{equation}
or
\begin{equation}
\label{eq:bicritcond1}
\beta = \alpha D,
\end{equation}

The first possibility, (\ref{eq:bicritcond2}), describes a line in $\alpha$-$\beta$ space, but the $\beta$-intercept $-aD + d$ is negative. Since $\alpha, \beta > 0$, the condition is realizable only for
\begin{equation}
\label{eq:betacond}
\alpha > \frac{aD-d}{D}.
\end{equation}
Substituting (\ref{eq:bicritcond2}), the wave number $q_{2,c}$ from (\ref{eq:q21}) is
\begin{equation}
q_{2,c} = \sqrt{\frac{3aD - d - 4\alpha D}{2D}}.
\end{equation}
For a Turing bifurcation, $q_{2,c}$ must be positive. Solving $q_{2,c} > 0$ and (\ref{eq:betacond}) simultaneously leads to the inequality $a < 5d/D$ which cannot be satisfied because of (\ref{eq:turingconds}). Hence, no Turing-Turing bifurcation is possible for (\ref{eq:bicritcond2}).

The second case, (\ref{eq:bicritcond1}), also describes a line in $\alpha$-$\beta$ space, but it emanates from the origin. Along this line, the wave number $q_{2,c}$ is
\begin{equation}
q_{2,c} = \sqrt{\frac{aD + d - 4\alpha D}{2D}},
\end{equation}
which is positive so long as
\begin{equation}
\label{eq:alpharange}
\alpha < \frac{aD+d}{4D}.
\end{equation}
Thus, for $0 < \alpha < (aD+d)/4D$ and $\beta = \alpha D$, codimension-two Turing-Turing bifurcations occur. The wave number ratio $r_q$ along this bifurcation curve is
\begin{equation}
\label{eq:rq}
r_q \equiv \frac{q_{1,c}}{q_{2,c}} = \sqrt{\frac{aD+d}{aD+d-4\alpha D}}.
\end{equation}
We will later use this result to choose chemical parameters giving rise to patterns dominated by a desired wave number (or alternatively, length scale) ratio.  In an experiment, changing the coupling for two chemical species independently is generally not possible, and hence novel experimental approaches would be needed to fulfill condition~(\ref{eq:bicritcond1}).

%%%%%%%%%%%%%%%%%%%%%%%%%%%%%%%%%%%%%%
\begin{figure}
\resizebox{\columnwidth}{!}{\includegraphics{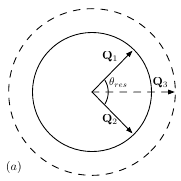} \includegraphics{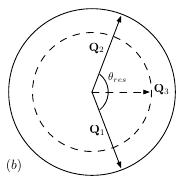}}
\caption{Diagram of resonant triads in Fourier space. The Fourier modes satisfy (\ref{eq:resonance}). Solid circles and vectors indicate neutral stability, and dotted ones indicate weak damping. In (a), $|\vec{Q}_{1,2}| < |\vec{Q}_3|$ and the resonant angle satisfies $0 \leq \theta_{res} < 2\pi/3$. In (b), $|\vec{Q}_{1,2}| > |\vec{Q}_3|$ and $2\pi/3 < \theta_{res} < \pi$. \label{fig:resonanttriads} }
\end{figure}
%%%%%%%%%%%%%%%%%%%%%%%%%%%%%%%%%%%%%%

The issue of wave number ratios connects to resonant triad interactions, which are important to the study of some pattern selection problems. Our discussion here echoes in some respects the discussions of  \cite{SilTopSke2000,TopSil2002,PorTopSil2004,TopPorSil2004}, which study resonant triads in Faraday waves. Resonant triad interactions, the lowest order nonlinear interactions, involve three modes with wave vectors $\vec{Q}_1$, $\vec{Q}_2$, and $\vec{Q}_3$ satisfying the condition
\begin{equation}
\label{eq:resonance}
\vec{Q}_1 + \vec{Q}_2 = \vec{Q}_3.
\end{equation}
For the resonant triads that interest us, $\vec{Q}_{1,2}$ lie on a single critical circle in Fourier space, and $\vec{Q}_3$ is a weakly damped mode lying on a different, (nearly) critical circle. Eq. (\ref{eq:resonance}) determines an angle of resonance $\theta_{res} \in [0,\pi)$ between the two critical wave vectors via the trigonometric relationship
\begin{equation}
\label{eq:thetares}
\cos \left( \frac{\theta_{res}}{2} \right) = \frac{Q_3}{2 Q_1},
\end{equation}
where $|\vec{Q}_1|=|\vec{Q}_2|=Q_1$ and $|\vec{Q}_3|=Q_3$. If $Q_1 < Q_3$ then $\theta_{res} \in [0,2\pi/3)$. If $Q_1 > Q_3$ then $\theta_{res} \in (2\pi/3,\pi)$. These two cases are pictured in Fig. \ref{fig:resonanttriads}.

Resonant triad interactions may impact pattern selection. Heuristically, the interaction allows energy exchange between the critical and damped modes. If the damped mode is a sink, drawing energy from the excited modes, the interaction is an anti-selection mechanism that suppresses patterns involving the resonant angle. Alternatively, if the damped mode is a source, feeding energy to the excited modes, patterns involving the resonant angle -- or equivalently, the associated length scale ratio -- may be enhanced.

For our reaction-diffusion system near the Turing-Turing bifurcation point, define $\lambda_1$ as the eigenvalue associated with $q_{1,c}$ having the largest real part; similarly for $\lambda_2$ and $q_{2,c}$. Now detune slightly in parameter space from the Turing-Turing bifurcation, so that $\lambda_{1,2}$ are small and oppositely signed. Consider the two different possibilities for resonant triads pictured in Fig.~\ref{fig:resonanttriads}. First, assume that the critical modes have a smaller wave number than the weakly damped one, so that panel (a) applies. Recalling that $q_{2,c} < q_{1,c}$ for the Turing-Turing bifurcation (we exclude the degenerate case of equality), this means that $Q_1 = q_{2,c}$ and $Q_3 = q_{1,c}$. Combining (\ref{eq:rq}) and  (\ref{eq:thetares}) gives the resonance angle at the Turing-Turing point,
\begin{equation}
\label{eq:res1}
\cos \left(\frac{\theta_{res}}{2} \right) = \frac{1}{2} \sqrt{\frac{aD+d}{aD+d-4\alpha D}}.
\end{equation}
The right-hand side must be real and must not exceed unit magnitude. These requirements yield an admissible range of $\alpha$ in which our resonant triads exist,
\begin{equation}
\label{eq:alphacond1}
0 < \alpha < \frac{3}{16} \frac{aD+d}{D},
\end{equation}
which is a subset of the range in (\ref{eq:alpharange}).

For the alternate case in which the critical modes have a larger wave number than the weakly damped one, Fig.~\ref{fig:resonanttriads}(b) applies. Then $Q_1 = q_{1,c}$ and $Q_3 = q_{2,c}$, and the resonance angle is
\begin{equation}
\label{eq:res2}
\cos \left(\frac{\theta_{res}}{2} \right) = \frac{1}{2} \sqrt{\frac{aD+d-4\alpha D}{aD+d}}.
\end{equation}
For this case, the entire range (\ref{eq:alpharange}) is admissible.

%To summarize, if a single two-component reaction-diffusion layer is poised at a Turing bifurcation, then introducing a coupled layer and enforcing (\ref{eq:bicritcond1}) yields a Turing-Turing bifurcation. Resonant triads involving wave numbers $q_{1,min}$ and $q_{2,min}$ exist nearby in parameter space. If (\ref{eq:alphacond1}) holds, there exist resonant triads for which the critical modes and damped mode have (approximately) wave numbers $q_{1,min}$ and $q_{2,min}$ respectively. The resonant angle lies in the range $0 < \theta_{res} < 2\pi/3$ as described by (\ref{eq:res1}).  If (\ref{eq:alphacond2}) holds, there exist resonant triads for which the critical modes and damped mode have (approximately) wave numbers $q_{2,min}$ and $q_{1,min}$ respectively. The resonant angle lies in the range $2\pi/3 < \theta_{res} < \pi$ as described by (\ref{eq:res2}).

\subsection{Multiple length scales in coupled Brusselator layers}
\label{sec:brusselator}

%%%%%%%%%%%%%%%%%%%%%%%%%%%%%%%%%%%%%%
\begin{figure}
\includegraphics{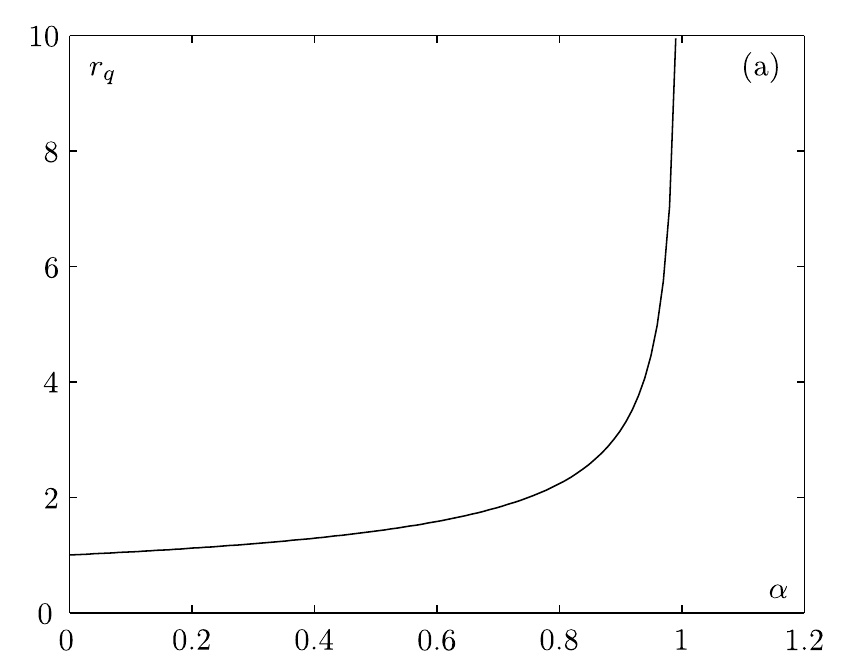}
\includegraphics{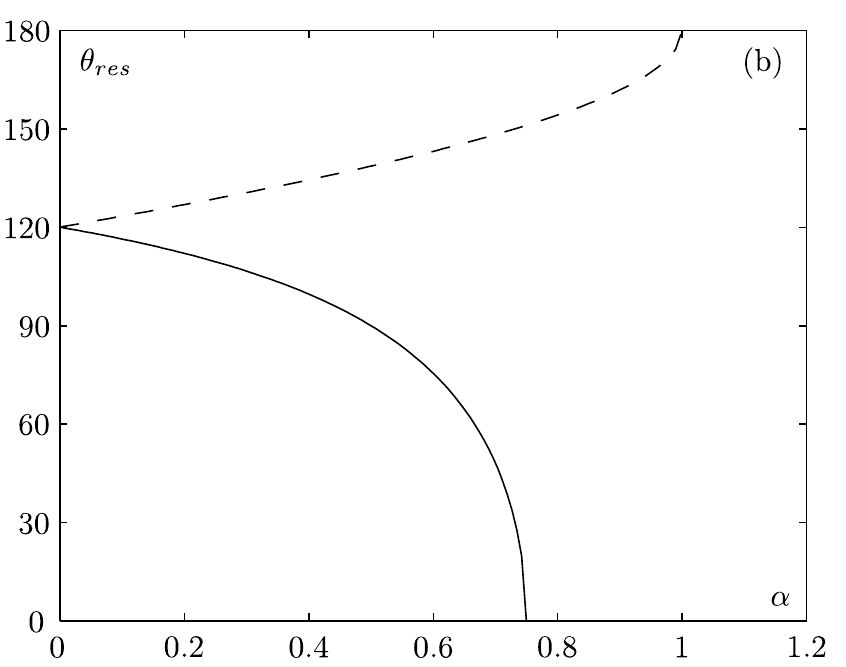}
\caption{(a) Ratio $r_q$ in (\ref{eq:rq}) of two (nearly) critical wave numbers near a Turing-Turing bifurcation in the Brusselator. The governing equations are (\ref{eq:twolayer}) and (\ref{eq:brusselator}) with $A=3$, $B=9$, $D = 2.25$. The coupling parameter $\alpha$ is a free parameter and $\beta = \alpha D$. (b) Angle of triad resonance corresponding to (a). For the lower (dashed) branch, the resonant triad corresponds to Fig. \ref{fig:resonanttriads}(a), in which the damped mode has larger wave number than the critical ones. For the upper (dashed) branch, the resonant triad corresponds to Fig. \ref{fig:resonanttriads}(b), in which the damped mode has smaller wave number. See Sec. \ref{sec:brusselator} for details. \label{fig:thetares}}
\end{figure}
%%%%%%%%%%%%%%%%%%%%%%%%%%%%%%%%%%%%%%

%%%%%%%%%%%%%%%%%%%%%%%%%%%%%%%%%%%%%%
\begin{figure*}
\centerline{\includegraphics{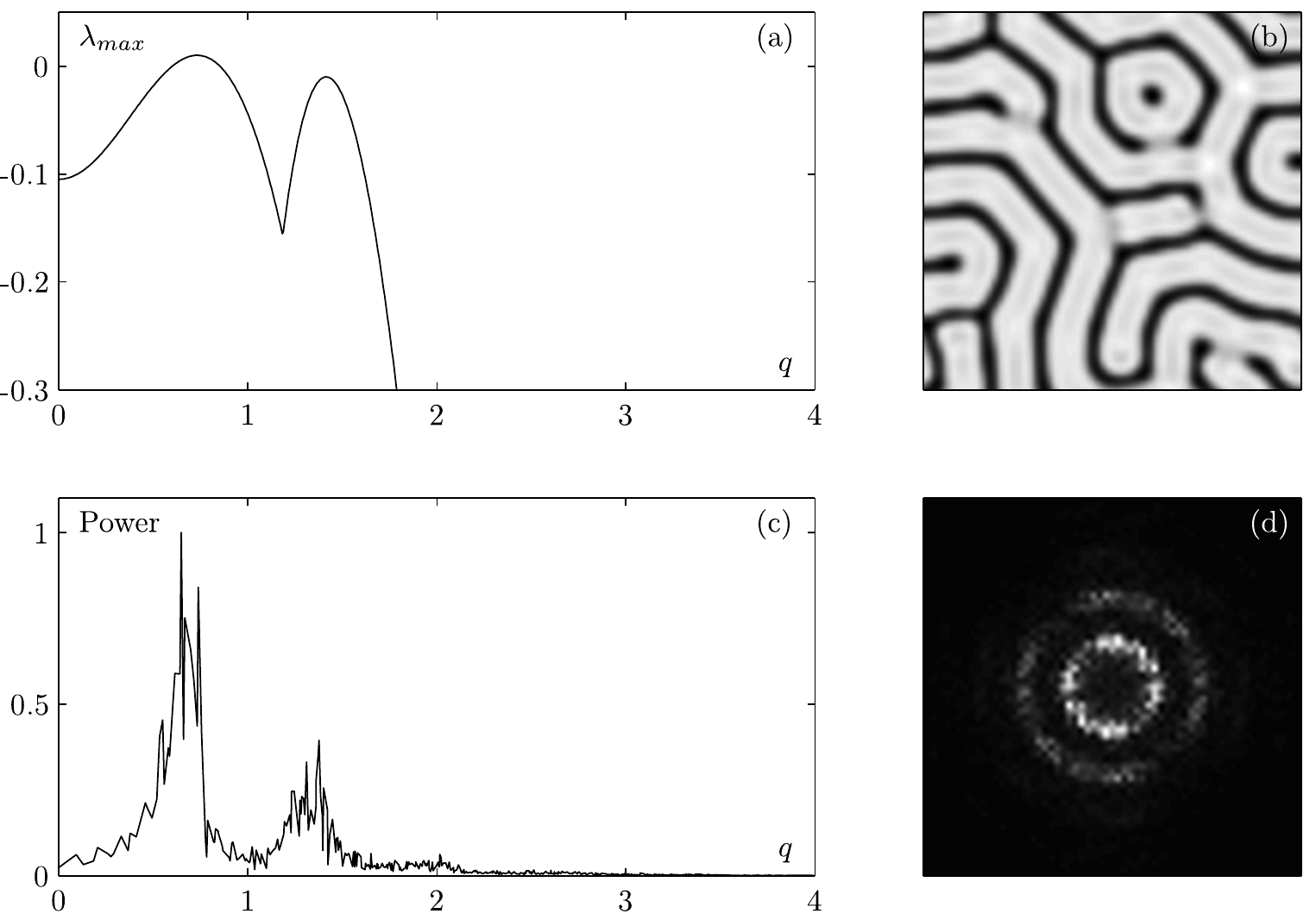}}
\caption{Numerical simulation of coupled Brusselators given by (\ref{eq:twolayer}) with (\ref{eq:brusselator}).  Parameter values are  $A=3$, $B=9$, $D = 2.244$, $\alpha = 0.723$, and $\beta = 1.633$ (to three decimal places).  (a) The eigenvalue with maximum real part is plotted as a function of wave number.  (b) Striped pattern resulting from this choice of parameters.  Dark and light regions indicated variations in concentration of chemical $u$ in the top layer.  The bottom layer looks the same but with light and dark regions reversed. (c) The radial power spectrum of the striped pattern with units chosen so that the dominant peak is normalized to unity.  (d) The Fourier spectrum of the striped pattern.    \label{fig:stripes} }
\end{figure*}
%%%%%%%%%%%%%%%%%%%%%%%%%%%%%%%%%%%%%%

As an example, we apply our results to the Brusselator~\cite{PriLef1968}. For this chemical reaction,
\begin{subequations}
\label{eq:brusselator}
\begin{eqnarray}
F(U,V) & = & A - (B+1) U + U^2 V,\\
G(U,V) & = & BU - U^2 V,
\end{eqnarray}
\end{subequations}
in (\ref{eq:twolayer}). $A,B$ are chemical parameters. The steady state is $(U^*,V^*)=(A,B/A)$ and the coefficients $a,b,c,d$ in (\ref{eq:linearops}) are
\begin{equation}
a = B - 1,\quad b = A^2,\quad c = -B,\quad d = -A^2.
\end{equation}
For concreteness, take $A=3$, $B=9$, as do many of the examples in \cite{YanDolZha2002}. Then
\begin{equation}
a = 8,\quad b = 9,\quad c = -9,\quad d = -9.
\end{equation}
To have $\Delta_{1,min} = 0$ in (\ref{eq:q11}), the diffusion coefficient must be $D = 2.25$. Then
\begin{equation}
q_{1,c}  \approx 1.414. 
\end{equation}
To have a codimension two bifurcation that admits resonant triads, (\ref{eq:bicritcond1}) must hold. Then
\begin{equation}
q_{2,c} = \sqrt{2 - 2 \alpha}.
\end{equation}

For these chemical parameters, Fig.~\ref{fig:thetares}(a) shows the wave number ratio $r_q$ in (\ref{eq:rq}) as a function of $\alpha$ at the Turing-Turing point. Fig.~\ref{fig:thetares}(b) shows the resonant triad angle $\theta_{res}$ in (\ref{eq:res1}) and (\ref{eq:res2}), also as a function of $\alpha$. For the lower (solid) branch, the resonant triad corresponds to Fig. \ref{fig:resonanttriads}(a), in which the damped mode has larger wave number than the critical ones. For the upper (dashed) branch, the resonant triad corresponds to Fig. \ref{fig:resonanttriads}(b), in which the damped mode has smaller wave number. For a range of $\alpha$, either branch is accessible, depending on how one detunes from the codimension-two point, \textit{i.e.}, which circle in Fourier space is damped.

\subsection{Numerical simulation}
\label{sec:numerics}

Using the linear stability results and the understanding of multiple critical length scales near the Turing-Turing bifurcation, we attempt to engineer patterns with desired ratios near the Turing-Turing point. As in Sec. \ref{sec:brusselator}, we adopt the Brusselator as our model and choose $A=3$, $B=9$ in (\ref{eq:brusselator}). We pre-select a desired wave length ratio and set parameters to be very near the Turing-Turing bifurcation, but such that one of the (nearly) critical modes has maximum eigenvalue of $0.01$ (and hence can grow) and the other (nearly) critical mode has maximum eigenvalue $-0.01$ (and hence is weakly damped). These conditions determine values of $D$, $\alpha$, and $\beta$. The computational domain is periodic and square, with the length of each side eight times the wave length of the weakly growing mode. Starting from a random initial condition, we integrate the system in spectral space with 64 modes along each axis using the {\tt E\sc{xpint}} exponential integrator package for {\tt M\sc{atlab}} \cite{BerSkaWri2007} with a time step of $h=0.4$ and Krogstad time-stepping. We run simulations to $t = 4000$, which for our parameter choices is long enough for the solution to approach an attractor.

%%%%%%%%%%%%%%%%%%%%%%%%%%%%%%%%%%%%%%
\begin{figure*}
\centerline{\includegraphics{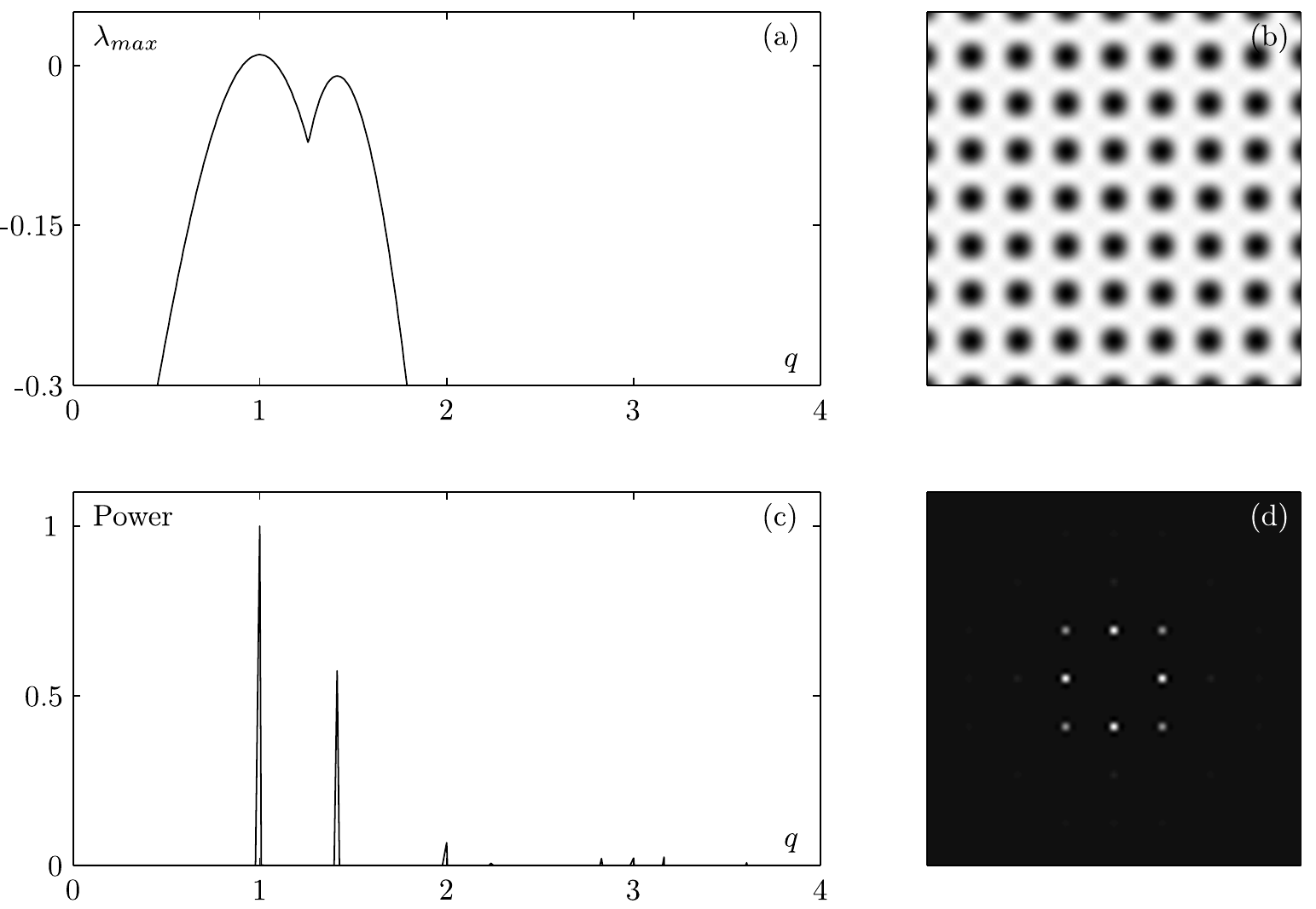}}
\caption{Parameter values are  $A=3$, $B=9$, $D = 2.244$, $\alpha = 0.490$, and $\beta = 1.113$ (to three decimal places).  (a) The eigenvalue with maximum real part is plotted as a function of wave number.  (b) Square pattern resulting from this choice of parameters.  Dark and light regions indicated variations in concentration of chemical $u$ in the top layer.  The bottom layer looks the same but with light and dark regions reversed. (c) The radial power spectrum of the square pattern with units chosen so that the dominant peak is normalized to unity.  (d) The Fourier spectrum of the square pattern.   \label{fig:squares} }
\end{figure*}
%%%%%%%%%%%%%%%%%%%%%%%%%%%%%%%%%%%%%%

In our first example, we select the wave number ratio  $0.5 \sec(\pi/12)$, corresponding to a resonant angle of $30\degree$.  These conditions determine $D = 2.244$, $\alpha = 0.723$, and $\beta = 1.633$ (to three decimal places). Thus the damped mode has wave number $q = 1.414$, and the dominant mode has wave number $q = 0.732$.  Fig. \ref{fig:stripes}(a) visualizes this, showing the (analytically calculated) eigenvalue with largest real part as a function of $q$. Fig. \ref{fig:stripes}(b) shows the result of the full numerical simulation, namely a stripe-dominated pattern that is sometimes referred to as labyrinthine. The Fourier spectrum of this pattern in  Fig. \ref{fig:stripes}(d) shows active modes lying on two circles in Fourier space (though it is clearly not dominated by resonant triad interactions). From the radial power spectrum of the pattern in Fig. \ref{fig:stripes}(c) (with units chosen so that the dominant peak is normalized to unity) we see that those circles correspond to the selected wave numbers.

A more ambitious goal is to go beyond selecting a ratio of length scales and to actually select a particular pattern. In general, this requires nonlinear analysis. However, we can show one example where harnessing the linear stability results does lead to successful pattern selection. For this case, we set the wave number ratio $\sqrt{2}$:1 so that the resonant angle is $90\degree$.  Optimistically, one might expect a square pattern, which is what we obtain in Fig. \ref{fig:squares}(b).  Fig. \ref{fig:squares}(d) shows that the angles between each dominant Fourier mode are $90^{\degree}$.  The chemical parameters in this case are the same as the previous example, except that we have changed the coupling parameters to $\alpha = 0.490$ and $\beta = 1.113$ in order to shift the (nearly) critical peak to the required value of $q_{1,c} \approx 1$, shown in Figs. \ref{fig:squares}(a) and (c). Steady square patterns have been reported in photosensitive reaction diffusion systems forced with a square mask \cite{BerYanDol2003}, and oscillatory square patterns have been observed in autonomous reaction-diffusion systems with interacting Turing and Hopf modes \cite{YanZhaEps2004}. We have not previously seen an unforced, steady square pattern reported in the chemical Turing pattern literature, and believe that our computational result in Fig.~\ref{fig:squares} represents the first such example.

\begin{figure*}
\centerline{\includegraphics{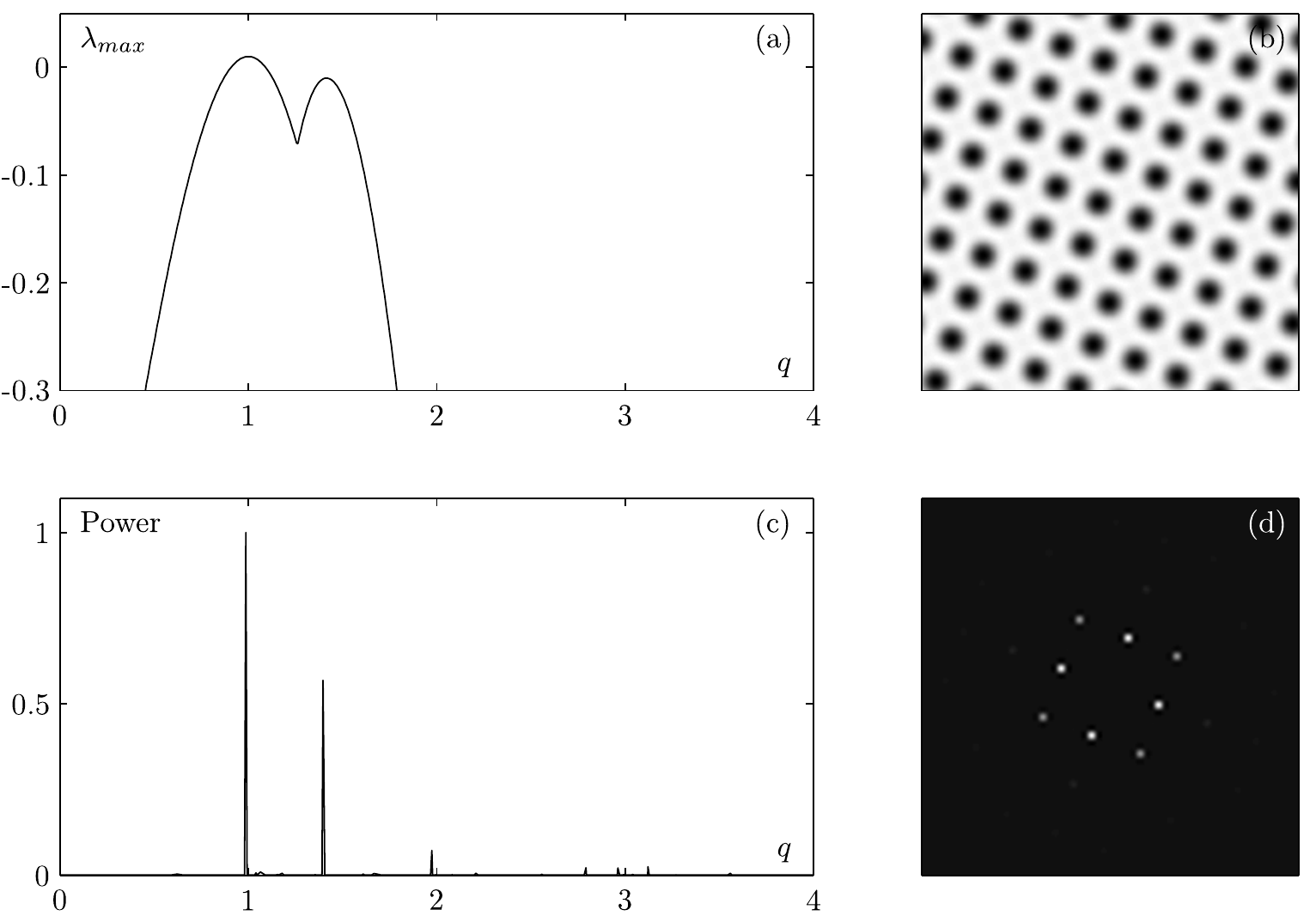}}
\caption{Results analogous to Fig.~\ref{fig:squares}, and with the same parameters. However, whereas the square computational domain in Fig.~\ref{fig:squares} had each side of length eight times the length of the weakly growing mode as determined from linear stability analysis, here we choose $5 \sqrt{3} \approx 8.7$ wavelengths per side in order to verify that the box size was not responsible for stabilizing the square pattern. Indeed, here we still obtain a square pattern, albeit one with a different orientation. (a) The eigenvalue with maximum real part is plotted as a function of wave number (identical to Fig.~\ref{fig:squares}(a), reproduced here for convenience).  (b) Square pattern.  Dark and light regions indicated variations in concentration of chemical $u$ in the top layer. (c) The radial power spectrum of the square pattern with units chosen so that the dominant peak is normalized to unity.  (d) The Fourier spectrum of the square pattern.  \label{fig:newsquares} }
\end{figure*}

To verify that the square pattern is robust to changes in domain size -- and not dependent on having a computational domain whose side fits an integral number of wavelengths of the weakly growing mode -- we repeat  the calculation of Fig.~\ref{fig:squares} but use a box size of $5 \sqrt{3} \approx 8.7$ wavelengths per side rather than eight, as before. This computation indeed still produces a square pattern, as shown in Fig.~\ref{fig:newsquares}, albeit one with a different spatial orientation.

%%%%%%%%%%%%%%%%%%%%%%%%%%%%%%%%%%%%%%
%%%%%%%%%%%%%%%%%%%%%%%%%%%%%%%%%%%%%%

\section{Conclusion}
\label{sec:conclusion}

Layered, spatially-extended reaction-diffusion systems are analytically taxing due to their (potentially high) dimensionality. The intriguing laboratory experiments and numerical simulations of the past decade have been supported by comparatively few theoretical works. In this paper, we have sought to develop some basic theory for simple layered scenarios, and to connect linear results to nonlinear pattern formation.

First, we presented a linear stability analysis for certain layered reaction-diffusion systems.  For two-layer systems of identical two-component layers, we analyzed the stability of homogeneous steady states by exploiting the block symmetric structure of the linear problem. This analysis revealed eight possible primary bifurcation scenarios, including a Turing-Turing bifurcation involving two length scales whose ratio may be tuned via the inter-layer coupling.  For systems of $n$-component layers and non-identical layers, the linear problem's block form allowed approximate decomposition into lower-dimensional linear problems for sufficiently weak coupling.

We applied some results to a two-layer Brusselator system near the Turing-Turing bifurcation. We calculated the ratio of critical wave numbers as a function of the coupling parameter and harnessed the analytical results to pre-selected chemical and coupling parameters that should give rise to a particular ratio in a fully (weakly) nonlinear system. Numerical simulations indeed revealed patterns dominated by the chosen ratio. In one example, by pre-selecting a $\sqrt{2}$:1 ratio, we obtained (without external forcing of the system) a simple, steady square-lattice-based pattern. Our numerical simulations demonstrate potential applications of our results as a means of understanding and engineering the instabilities in layered reaction-diffusion systems. However, to develop a more complete picture of pattern formation in these systems, nonlinear analysis is required.  We expect future work could address these detailed questions of pattern selection.

Finally, we hope that our results might be of use to experimentalists.  For instance, the Lengyel-Epstein model of the two-layer CDIMA reaction \cite{YanEps2004} can be written with \mbox{$F(U,V)  =  A -U - 4UV/(1+U^2)$}, \mbox{$G(U,V) = BU - BUV(1+U^2)$} in~(\ref{eq:twolayer}).
%The steady state is $(U^*,V^*)=(A/5,1+A^2/25)$ and
The coefficients $a,b,c,d$ in (\ref{eq:linearops}) are $a = (3A^2-125)/\gamma$, $b = -20A/\gamma$, $c = 2 A^2B/\gamma$, and $d = -5 AB/\gamma$, where for convenience we define $\gamma = A^2+25$. Assuming that the Turing-Turing bifurcation conditions of Sec. \ref{sec:resonanttriads} are met, the wave number ratio $r_q$ in (\ref{eq:rq}) is $\sqrt{2}$:1 when
\begin{equation}
\alpha = \frac{3A^2D -5 A B - 125D}{8D(A^2+25)}.
\end{equation}
Experiments performed in this parameter regime might shed light on whether steady square-lattice-based patterns can indeed arise.

\begin{acknowledgments}
This work was supported by NSF grants DMS-0740484 and DMS-1009633. Portions of the research were completed by AM as part of her senior capstone project in mathematics at Macalester College. We are grateful to Tom Halverson for helpful conversations.
\end{acknowledgments}

\appendix

\section{Eigenvalues of block matrices}
\label{sec:block}

Here we show a useful identity for the eigenvalues of a block matrix with the symmetric form relevant to the stability calculation in Sec. \ref{sec:linearderiv} - \ref{sec:primbif}.

First we perform a side calculation. Consider a block matrix of the form
\begin{equation}
\label{eq:appendixL}
\mat{L} = \begin{pmatrix} \mat{P} & \mat{Q} \\ \mat{R} & \mat{S} \end{pmatrix}.
\end{equation}
Assume $\mat{S}$ is invertible and factor this as
\begin{equation}
\label{eq:blockfactor}
\mat{L} = \begin{pmatrix} \mat{I} & \mat{Q} \\ \mat{0} & \mat{S} \end{pmatrix} \begin{pmatrix} \mat{P-QS^{-1}R} & \mat{0} \\ \mat{S^{-1}R} & \mat{I} \end{pmatrix},
\end{equation}
where $\mat{I}$ is the (appropriately sized) identity matrix. Now apply results from \cite{Sil2000} for determinants of block matrices. For the factors in (\ref{eq:blockfactor}), we have
\begin{equation}
\label{eq:blockfactor1}
\det \begin{pmatrix} \mat{I} & \mat{Q} \\ \mat{0} & \mat{S} \end{pmatrix} = \det(\mat{I}) \det(\mat{S}) = \det(\mat{S}),\\
\end{equation}
and
\begin{subequations}
\label{eq:blockfactor2}
\begin{eqnarray}
& \phantom{=} &\det \begin{pmatrix} \mat{P-QS^{-1}R} & \mat{0} \\ \mat{S^{-1}R} & \mat{I} \end{pmatrix}  \\
& = & \det(\mat{P-QS^{-1}R}) \det(\mat{I}) \\
& = & \det(\mat{P-QS^{-1}R}).
\end{eqnarray}
\end{subequations}
Combine (\ref{eq:blockfactor}) - (\ref{eq:blockfactor2}) to obtain
\begin{equation}
\label{eq:blockdet}
\det(\mat{L}) = \det (\mat{S}) \det (\mat{P} - \mat{Q} \mat{S}^{-1} \mat{R}).
\end{equation}

We now turn to our main calculation of this Appendix. Consider the stability analysis in Sec. \ref{sec:linearderiv} - \ref{sec:primbif}, in which case $\mat{R}=\mat{Q}$, $\mat{S}=\mat{P}$  in (\ref{eq:appendixL}) and $\mat{P}$ and $\mat{Q}$ are identically-sized square matrices. That is,
\begin{equation}
\label{eq:symmetricL}
\mat{L} = \begin{pmatrix} \mat{P} & \mat{Q} \\ \mat{Q} & \mat{P} \end{pmatrix}.
\end{equation}
Seek the eigenvalues by finding the roots of the characteristic polynomial $C_{\mat{L}}(\lambda) =  \det(\mat{L} - \lambda \mat{I})$, or more explicitly,
\begin{equation}
C_{\mat{L}}(\lambda) = \det \begin{pmatrix} \mat{P}-\lambda\mat{I} & \mat{Q} \\ \mat{Q} & \mat{P}-\lambda\mat{I} \end{pmatrix}.
\end{equation}
Then $C_{\mat{L}}(\lambda)$ takes the form
\begin{widetext}
\begin{subequations}
\begin{eqnarray}
C_{\mat{L}}(\lambda) & = & \det[\mat{P} - \lambda \mat{I}]^{\phantom{2}} \det [  \mat{P} - \lambda  \mat{I} - \mat{Q}(\mat{P} - \lambda  \mat{I})^{-1} \mat{Q}], \\
& = &  \det[\mat{P} - \lambda \mat{I}]^2 \det [\mat{I} -  (\mat{P} - \lambda  \mat{I})^{-1} \mat{Q}(\mat{P} - \lambda  \mat{I})^{-1} \mat{Q}], \\
& = &  \det[\mat{P} - \lambda \mat{I}]^2 \det [\mat{I} -  \{(\mat{P} - \lambda  \mat{I})^{-1} \mat{Q}\}^2], \\
& = &  \det[\mat{P} - \lambda \mat{I}]^2 \det [\mat{I} -  (\mat{P} - \lambda  \mat{I})^{-1} \mat{Q}] \det [\mat{I} +  (\mat{P} - \lambda  \mat{I})^{-1} \mat{Q}], \\
& = &  \det[\mat{P} - \lambda \mat{I} - \mat{Q}] \det[\mat{P} - \lambda \mat{I} + \mat{Q}], \\
& = &  \det[(\mat{P}-\mat{Q}) - \lambda \mat{I}] \det[(\mat{P}+\mat{Q}) - \lambda \mat{I}], \\
& = & C_{\mat{P-Q}}(\lambda) \cdot C_{\mat{P+Q}}(\lambda).
\end{eqnarray}
\end{subequations}
\end{widetext}
The first line follows from direct application of (\ref{eq:blockdet}). The second follows from pulling a factor of $\mat{P}-\lambda \mat{I}$ out of the second determinant and combining it with the first. The third line follows from noting the squared quantity. The fourth line follows from factoring a difference of squares. The fifth line follows from redistributing one factor of $\mat{P}-\lambda \mat{I}$ into each of the two other terms. The sixth line follows simply from commutativity of matrix addition/subtraction, and the last line follows from the definition of a characteristic polynomial.

Thus, the characteristic polynomial for (\ref{eq:symmetricL}) factors into that of $\mat{P} - \mat{Q}$ and $\mat{P} + \mat{Q}$, and therefore, the eigenvalues of $\mat{L}$ in (\ref{eq:symmetricL}) are the eigenvalues of  $\mat{P} - \mat{Q}$ and the eigenvalues of  $\mat{P} + \mat{Q}$.

\section{Eigenvalues of block matrices with blocks that are small in magnitude}
\label{sec:block2}
We now show an approximation for the eigenvalues for a block matrix of a particular form, where certain blocks are scaled by a small parameter. Begin with the matrix
\begin{equation}
\label{eq:nonsymmetricL}
\mat{L} = \begin{pmatrix} \mat{P} & \mat{Q} \\ \mat{Q} & \mat{S} \end{pmatrix},
\end{equation}
which arises as the linearization of a problem considered in Sec. \ref{sec:extensions}. In fact, $\mat{P}$ and $\mat{S}$ include additive factors of $\mat{Q}$, so for convenience, we let $\mat{P} = \widetilde{\mat{P}} - \mat{Q}$ and $\mat{S} = \widetilde{\mat{S}} - \mat{Q}$. For the case of weak chemical coupling, the entries in $\mat{Q}$ are small, so we let $\mat{Q} \to \epsilon \mat{Q}$ where $\epsilon \ll 1$ is a small bookkeeping parameter. Our matrix now has the form 
\begin{equation}
\mat{L} = \begin{pmatrix} \widetilde{\mat{P}} - \epsilon \mat{Q} & \epsilon \mat{Q} \\ \epsilon \mat{Q} & \widetilde{\mat{S}} - \epsilon \mat{Q} \end{pmatrix}.
\end{equation}
The characteristic polynomial is 
\begin{widetext}
\begin{eqnarray}
C_{\mat{L}}(\lambda) & = & \det[\widetilde{\mat{S}}-\epsilon \mat{Q} -  \lambda \mat{I}]  \det [\widetilde{\mat{P}}-\epsilon \mat{Q} -  \lambda \mat{I} - \epsilon^2 \mat{Q} (\widetilde{\mat{S}} - \epsilon \mat{Q} - \lambda \mat{I})^{-1} \mat{Q}] \label{eq:epsilon1} \\
& = & \det[\widetilde{\mat{S}}-\epsilon \mat{Q} -  \lambda \mat{I}]  \bigl\{  \det[\widetilde{\mat{P}}-\epsilon \mat{Q} -  \lambda \mat{I}]  + \order(\epsilon^2) \bigr\},\label{eq:epsilon2} \\	
& \approx & \det[\widetilde{\mat{S}}-\epsilon \mat{Q} -  \lambda \mat{I}] \det[\widetilde{\mat{P}}-\epsilon \mat{Q} -  \lambda \mat{I}], \label{eq:epsilon3} \\ 
& = & C_{\mat{S}}(\lambda) \cdot C_{\mat{P}}(\lambda). \label{eq:epsilon4}
\end{eqnarray}
\end{widetext}
The first line follows from direct application of (\ref{eq:blockdet}). The second line follows from Jacobi's formula for the differential of a determinant. The third line follows from neglecting the $\order(\epsilon^2)$ correction, and the final line follows from the definitions of $\mat{S}$ and $\mat{P}$, and from the definition of a characteristic polynomial. Thus, the eigenvalues of (\ref{eq:nonsymmetricL}) are approximately those of $\mat{P}$ and those of $\mat{S}$ so long as $\mat{Q}$ is scaled by a small parameter.

%\bibliography{/Users/chad/Dropbox/Professional/Research/library/master_bibliography}
\bibliography{master_bibliography}

\end{document}